\documentclass[aps,prb,twocolumn,showpacs,preprintnumbers,amsmath,amssymb,superscriptaddress,noshowkeys]{revtex4}
\usepackage{txfonts}
\usepackage[dvipdfm]{graphicx}
\usepackage{bm}
\def\rnum#1{\expandafter{\romannumeral #1}} 
\def\Rnum#1{\uppercase\expandafter{\romannumeral #1}}

\newfont{\bg}{cmr10 scaled\magstep4}

\newcommand{\bigzerou}{\smash{\lower1.8ex\hbox{\bg 0}}}

\begin{document}

\title{Singularities of Andreev spectrum in multi-terminal Josephson junction}

\author{Tomohiro Yokoyama}
\email[E-mail me at: ]{T.Yokoyama@tudelft.nl}
\affiliation{Kavli Institute of Nanoscience, Delft University of Technology,
Lorentzweg 1, 2628 CJ, Delft, The Netherlands}
\affiliation{Center for Emergent Matter Science, RIKEN institute,
2-1 Hirosawa, Wako, Saitama 351-0198, Japan}
\author{Yuli V.\ Nazarov}
\affiliation{Kavli Institute of Nanoscience, Delft University of Technology,
Lorentzweg 1, 2628 CJ, Delft, The Netherlands}

\date{\today}

\begin{abstract}
The energies of Andreev bound states (ABS)  forming in a $N$-terminal junction are affected by $N - 1$ independent
macroscopic phase differences between superconducting leads and can be regarded as energy bands in
$N - 1$ periodic solid owing to the $2\pi$ periodicity in all phases.
We investigate the singularities and peculiarities of the resulting ABS spectrum combining phenomenological and
analytical methods and illustrating with the numerical results. We pay special attention on spin-orbit (SO) effects.
We consider Weyl singularities with a conical spectrum that are situated at zero energy in the absence of SO interaction.
We show that the SO interaction splits the spectrum in spin like a Zeeman field would do.
The singularity is preserved while departed from zero energy. With SO interaction, points of zero-energy  form
an $N - 2$ dimensional manifold in $N - 1$ dimensional space of phases, while this dimension is $N - 3$ in
the absence of SO interaction.
The singularities of other type are situated near the superconducting gap edge. 
In the absence (presence) of SO interaction, the ABS spectrum at the gap edge is  mathematically analogues to
that at zero energy in the presence (absence) of SO interaction. We demonstrate that the gap edge
touching (GET) points of the spectrum in principle form $N - 2$ ($N - 3$) dimensional manifold when
the SO interaction is absent (present). Certain symmetry lines in the Brillouin zone of the phases are
exceptional from this rule, and GET there should be considered separately.
We derive and study the effective Hamiltonians for all the singularities under consideration.
\end{abstract}
%\preprint
\maketitle

\section{INTRODUCTION}

Superconducting junctions give rise to many interesting and unique physical phenomena,
this being a base of the numerous applications in the field of quantum devices.
A conventional Josephson junction with two superconducting leads hosts the Andreev bound states (ABS),
that carry the supercurrent determined by the difference of macroscopic phases of the leads~\cite{NazarovBlanter}.
The properties of ABS may be altered by connecting the superconductors with special materials.
For example, the exchange field in a ferromagnetic junction splits the ABS energies in spin.
This may result in the $\pi$-state~\cite{Oboznov}. Recent studies address topologically protected bound states with
zero-energy, called Majorana bound states, that occur in the 1D semiconductor nanowire with spin-orbit (SO) interaction,
Zeeman splitting, and proximity-induced superconducting gap~\cite{Mourik,Rokhinson,Das,Deng}.
The presence of Majorana bound states in the junction may double the period of current-phase
relation~\cite{FuKane,Lutchyn,Beenakker1}.
The coexistence of SO interaction and Zeeman effect breaks the spin-rotation and time-reversal symmetries.
With symmetry broken, the Josephson current is not an odd function of the phase difference.
This is called the anomalous Josephson effect~\cite{Buzdin,Reynoso1,Reynoso2,YEN1,YEN2}.

The Josephson junctions involving various materials have been mostly investigated in two-terminal setups.
There is a recent interest in multi-terminal Josephson junctions~\cite{Heck,Riwar,Padurariu}.
Such junctions have been realized, for instance, with crossed InSb/As nanowires~\cite{Plissard},
where SO interaction is strong. Multi-terminal Josephson junction with $N$ superconducting leads is
affected by $N-1$ independent phase differences. The energies of ABS are $2\pi$ periodic in all phase differences.
The system of energy levels of ABS can be regarded as a band structure in a $N-1$ dimensions.
The phase differences play the role of quasimomenta~\cite{Riwar}.
The multi-terminal junctions may exhibit topological properties even if the superconducting leads and
the connecting region are not made from topological or other exotic materials. In the case of two-terminal junction,
the Andreev levels touch zero energy only when the transmission coefficient of the normal region is unity and
the phase difference is $\varphi = \pm \pi$. For the multi-terminal junctions, the Andreev levels can reach
zero energy at some isolated points in $N-1$ dimensional space of phase differences~\cite{Heck,Riwar,Padurariu}.
Such points are topologically protected being the Weyl singularities studied theoretically in 3D solids~\cite{Wan}.

The energy gap closes at a Weyl point and satisfies a conical dispersion in the vicinity of the point.
The Berry curvature field is divergent at the Weyl points. They can be regarded as Dirac monopoles of
the Berry curvature field bearing the topological charge, $\pm 1$. A band structure with Weyl points can be
continuously transformed to that without the points if two Weyl points with opposite topological charge meet
each other to annihilate. In a 3D solid, the SO interaction and the inversion-symmetry breaking are
essential for occurrence stable Weyl point~\cite{Wan,Burkov,Murakami}.
Very recently, the angle-resolved photoemission spectroscopy experiments confirmed the existence of
the Weyl points in the 3D solids, such as TaAs~\cite{SuYXu1,Lv,Yang}, TaP~\cite{NXu}, and NbAs~\cite{SuYXu2}.

Riwar {\it et al.}~\cite{Riwar} demonstrate the presence of Weyl points in the multi-terminal short
Josephson junctions. In contrast to solids, the Weyl point does not require SO interaction.
The Andreev levels are Kramers-degenerate for the whole ($N-1$)-dimensional $\varphi$-space.
They discuss the transconductance to detect the Chern number due to the Weyl points by using
one phase as a control parameter to switch the topological state. To reveal experimental signatures of
the topology associated with Weyl points, the authors of Ref.\ \onlinecite{Riwar} propose the following scheme.
They consider a 2D band structure that depends on the two phases, $\varphi_1$ and $\varphi_2$.
The property of this band structure can be tuned by the remaining phase, $\varphi_3$.
The 2D band structure is characterized by a Chern number that is proportional to the flux of
Berry curvature field through a ($\varphi_1 ,\varphi_2$) 2D plane.
The Chern number is changed by one anytime the plane crosses the position of Weyl singularity.
The Chern number is observed as a quantized transconductance between the leads one and two,
in similarity with the quantum Hall effect.

Hech {\it et al.}~\cite{Heck} and Padurariu {\it et al.}~\cite{Padurariu} also study the Andreev levels at
energies close to zero in three-terminal Josephson junction. The authors claim that the zero energy
states in such junctions may open opportunity for a single fermion manipulation. In the three-terminal junctions,
the Weyl singularity is generally absent although the energy of the ABS can pass zero.
The authors obtain a condition for zero energy ABS and study the density of states in detail.
When SO interaction is present, the energy levels of the ABS are split in spin.

In this study, we investigate theoretically the singularities of the ABS spectrum in four-terminal Josephson
junction taking SO interaction into account. We thus attempt to formulate the full picture of such singularities
combining phenomenological and analytical methods and illustrating it with numerical results.
The ABS energies are found from the Beenakker's determinant equation~\cite{Beenakker2}
using the scattering matrix of the junction. Mostly we concentrate on the case of short junction where
we can disregard the energy dependence of scattering matrix. Sometimes the absence of energy dependence
leads to extra degeneracy in the spectrum. To lift those we take into account the energy dependence by
perturbation theory.

Firstly, we concentrate on Weyl singularities that occur at $E = 0$ in the absence of SO interaction.
We consider the behavior of the singularities upon gradual increase of the strength of SO interaction.
We have found that SO interaction splits the spectrum of ABS in spin. Very much like Zeeman effect would do.
The conical points are departed from $E = 0$ to mirror symmetric positive and negative energies.
The Weyl singularities thus remain topologically protected. A small modification of scattering matrix by a parameter
changes the position of the mirror symmetric conical points, not eliminating them. As we show in numerical illustration,
an annihilation of Weyl points of opposite charge can take place upon the tuning of the scattering matrix.
We derive the effective Hamiltonian describing the vicinity of the Weyl points. The cones in the vicinity intersect
$E = 0$ at a 2D surface in a 3D space of the phases. We prove that this is the general property of ABS spectrum in
the presence of SO interaction.

Singularities of a different type arise when ABS energies approaches the gap edge, $E = \Delta$.
Owing to mirror symmetry of Andreev spectrum, there is a level with $E = -\Delta$ at the same position in
the space of the phases. We formulate a mathematical analogy that permits to map Weyl singularities at
$E = 0$ in the presence (absence) of the SO interaction to singularities at
$E = \Delta$ in the absence (presence) of the SO interaction. Since the ABS energies in the presence of
SO interaction reach zero energy at 2D surface, we expect the gap touching point to form 2D surfaces in
the absence of SO interaction. Indeed this can be seen in a concrete numerical calculation.
Employing the same analogy from the fact that without SO interaction the ABS energies reach zero at
isolated point only, we derive that the gap edge touching (GET) in the presence of SO interaction generally occurs
only at the isolated point. This implies that even a weak SO interaction removes the GET.
We construct the effective Hamiltonian to describe this situation. In this case, a weak energy dependence of
scattering matrix can also become important. An important peculiarity of the ABS spectrum concerns the vicinity of
symmetry lines in 3D elementary cell of the space of the phases. Three of the four superconducting phases are
the same at a symmetry line. Therefore the four-terminal junction at a symmetry line can be regarded as
a two-terminal junction with unequal number of conduction channels in the two leads.
As mentioned in Ref.\ \onlinecite{Beri}, for two-terminal short junction, the SO interaction is irrelevant to
causing spin splitting. Thus the vicinity of symmetry lines requires a separate consideration.
We derive an effective Hamiltonian to incorporate the details of the GET in the vicinity of the symmetry lines.

These two types of singular point would reveal the topological nature of multi-terminal Josephson junctions.
It brings a goal to propose nanostructures as artificial exotic materials.

The structure of this article is as follows.
In Sec.\ II, we explain the model for the multi-terminal Josephson junction and
the equation to determine the ABS energies.
In Sec.\ III, we describe the spin splitting of Weyl singularities.
Section IV is devoted to the GET point in general.
Here we formulate and employ the mapping between $E=0$ and $E=\Delta$.
We also concentrate separately at the vicinity of the symmetry lines in this Section.
We conclude in the last Section.

\section{MODEL AND FORMULATION}
\label{sec:model}

In this Section, we explain the model in use.

\begin{figure}
\includegraphics[width=50mm]{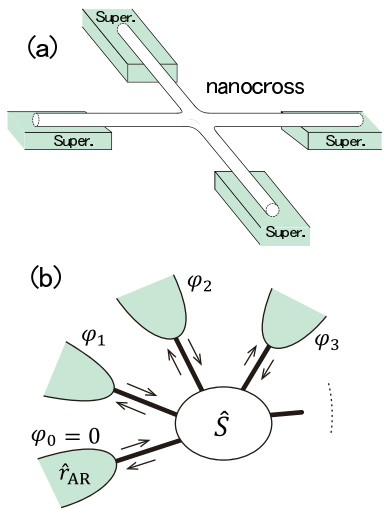}
\caption{
(a) Schematics of a four-terminal junction based on crossing semiconductor nanowires,
(nanocross)~\cite{Plissard}.
(b) Scattering model for multi-terminal Josephson junction. The junction itself is described by
the scattering matrix $\hat{S}$. The superconducting leads provide Andreev reflection with
amplitude $\hat{r}_{\rm AR}$. We assume the same order parameter in all superconducting leads
while the phases $\varphi_{0,1,\cdots,N-1}$ can differ. By virtue of gauge invariance,
one of the phases can be conveniently set to zero.
}
\label{fig:model}
\end{figure}

We consider a junction connected to $N$ superconducting leads. An example of a physical
system of this sort is given in a Fig.\ \ref{fig:model}(a).
All microscopic detail of the junction can be incorporated into the scattering matrix $\hat{s}_{\rm e,h}$ of
the electrons and holes (Fig.\ \ref{fig:model}(b)). We assume certain numbers $N_i$ of spin-degenerate
transport channels in each lead, so $\hat{s}_{\rm e,h}$ are $2M \times 2M$ matrices, $M = \sum_{i=0}^{N-1} N_i$.
The symmetry of Bogoliubov de-Gennes equation implies the relation $\hat{s}_{\rm h} (E) = - \hat{g} \hat{s}_{\rm e}^* (-E) \hat{g}$
with $\hat{g} = -i \hat{\sigma}_y$ being a matrix realizing a time-inversion in the spin space. In addition to this,
the requirement of time reversibility implies $\hat{s}_{\rm e} (E) = - \hat{g} \hat{s}_{\rm e}^{\rm T} (E) \hat{g}$.
The superconducting leads do not provide extra potential scattering. They are described by the Andreev reflection
amplitudes for converting electron to hole and hole to electron that do not change transport channel index
and are presented by diagonal matrix $\hat{r}_{\rm he, eh} = e^{\mp i\hat{\varphi}} e^{-i \hat{\chi}}$.
$\varphi_k$ being a superconducting phase of a lead corresponding to the channel $k$.
$\chi_k$ is an energy-dependent phase of Andreev reflection.
We assume the same material for all of the superconducting leads, so that their order parameters are the same
$\Delta_i = \Delta$. In this case, the energy-dependent phase is the same for the all leads and is given by
$\chi = \arccos (E/\Delta)$. The eigenvectors of the ABS satisfy
$\vec{\psi} = \hat{r}_{\rm eh} \hat{s}_{\rm h} \hat{r}_{\rm he} \hat{s}_{\rm e} \vec{\psi}$.
Therefore the energies of ABS are determined from~\cite{Beenakker2}
\begin{equation}
\det \left( e^{i 2\chi} - S(\vec{\varphi},E) \right) =0
\label{eq:Beenakker}
\end{equation}
with
\begin{eqnarray}
S (\vec{\varphi},E) &=& -\hat{g} s^* (\vec{\varphi},-E) \hat{g} s (\vec{\varphi},E),
\label{eq:Smatrix} \\
s (\vec{\varphi},E) &\equiv & e^{-i \hat{\varphi}/2} \hat{s}_{\rm e} (E) e^{+i \hat{\varphi}/2}.
\label{eq:smalls}
\end{eqnarray}
The hat symbol on $s$ and $S$ is omitted for simplification. Note that $S (\vec{\varphi},E)$ is a unitary matrix.
%although the time-reversal symmetry is broken by the phase differences,
%$s (\vec{\varphi},E) \ne -\hat{g} s (\vec{\varphi},E)^{\rm T} \hat{g}$.

For numerical calculations, the scattering matrix $\hat{s}_{\rm e}$ is taken as a random matrix.
The SO interaction is taken into account in $\hat{s}_{\rm e}$. In the presence (absence) of
SO interaction, random matrix is a member of symplectic (orthogonal) ensemble.
We introduce a parameter $p_{\rm SO}$ tuning the strength of SO interaction~\cite{YEN1}.
The parameter varies from $0$ to $1$, providing a continuous transition between
the orthogonal ($p_{\rm SO} =0$) and the symplectic ($p_{\rm SO} =1$) ensembles.

Let us discuss the ABS energies in the vicinities of zero energy, $E = 0$, and
superconducting gap edge, $E = \Delta$, where the first term in the determinant in
Eq.\ (\ref{eq:Beenakker}) becomes $e^{i 2\chi} = -1$ and $+1$, respectively.
In the absence of SO interaction, the scattering matrix $s (\vec{\varphi},E)$ commutes with
$\hat{g}$. Thus,
\begin{equation}
S (\vec{\varphi},E) = s^* (\vec{\varphi},-E) s (\vec{\varphi},E).
\label{eq:ss}
\end{equation}
When the SO interaction is present, we can introduce a unitary matrix
$u(\vec{\varphi},E) \equiv \hat{g} s(\vec{\varphi},E)$. Since $\hat{g}$ is a real matrix,
\begin{equation}
S (\vec{\varphi},E) = - u^* (\vec{\varphi},-E) u (\vec{\varphi},E).
\label{eq:uu}
\end{equation}
Let us disregard energy dependence of scattering matrices. Suppose that for
a sufficiently general unitary matrix $v$, there is eigenvector $|\psi \rangle = v^* v |\psi \rangle$.
Such eigenvector would correspond to an ABS at $E = 0$ in the presence of SO interaction and
to an ABS at $E = \Delta$ in the absence of SO interaction. Correspondingly, an eigenvector
$-|\psi \rangle = v^* v |\psi \rangle$ would be an ABS at $E = 0$ in the absence of SO interaction
and to an ABS at $E = \Delta$ in the presence of SO interaction.
We thus establish a mapping between the situation at $E = \Delta$ ($E=0$) with SO interaction
and at $E = 0$ ($E = \Delta$) without SO interaction.

\section{WEYL SINGULARITIES AT $E \approx 0$}
\label{sec:Weyl}

In this Section, we concentrate on the Weyl singularities. In our numerical illustrations,
we concentrate on four-terminal short junctions with a single channel in each lead, $N_i = 1$ and $M = 4$.
In this case, the ABS energies are periodic function of three independent phases
$\vec{\varphi} \equiv (\varphi_1, \varphi_2, \varphi_3)$ so we can restrict ourselves to
the Brillouin zone $|\varphi_i| \leq \pi$ for $i=1,2,3$. The time-reversibility manifests itself as
the inversion symmetry in this Brillouin zone, $E(\vec{\varphi}) = E(-\vec{\varphi})$.
To start with, we demonstrate that in the absence of SO interaction the energy levels of ABS exhibit
the band gap closing points at zero energy~\cite{Riwar}.
Next, we continuously change the scattering matrix increasing the parameter $p_{\rm SO}$,
that is strength of SO interaction. We demonstrate that the SO interaction splits the conical spectrum in spin.
The conical point departs from $E = 0$ and the energy levels cross zero energy at
2D surface rather than isolated point. This proves topological protection of the Weyl singularity.
To prove the generality of our conclusions, we derive an effective Hamiltonian from Eq.\ (\ref{eq:Beenakker})
that is valid in  the vicinity of the singularity and at weak SO interaction.

\begin{figure}
\includegraphics[width=85mm]{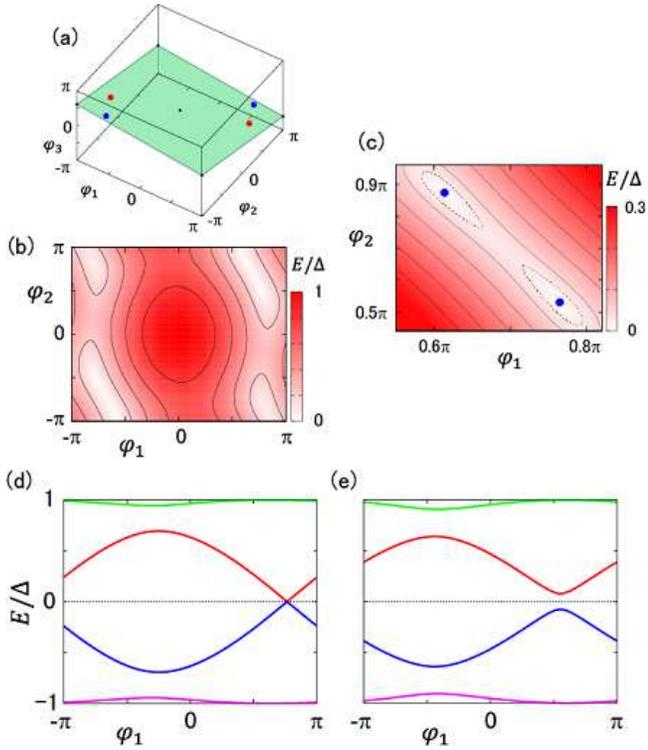}
\caption{(Color online)
Energies of the ABS in the four-terminal junction for a particular choice of $\hat{s}_{\rm e}$.
The SO interaction is absent ($p_{\rm SO} = 0$). For this choice, there are four Weyl singularities at $E = 0$.
(a) The positions of the Weyl singularities in the 3D space of the phases.
All four points and the origin lie in the same 2D plane.
(b) The gray scale plot of the lowest positive ABS energy in the 2D plane shown in (a).
Contour lines indicate $E/\Delta = 0.2$, $0.4$, $0.6$, and $0.8$.
(c) A zoom of the energy landscape (b) in the vicinity of Weyl singularities.
The dots give the singularity positions. The dashed contour line indicates $E/\Delta = 0.025$.
The solid contour lines corresponds to $E/\Delta$ in multiples of $0.05$.
Panels (d) and (e) show the ABS energies when two of the three phases are fixed to
(d) $\varphi_2 \simeq 0.53\pi$ and $\varphi_3 \simeq -0.42\pi$ and
(e) $\varphi_2 \simeq 0.53\pi$ and $\varphi_3 = -0.60\pi$.
}
\label{fig:E0noSO}
\end{figure}

\subsection{Energies of the Andreev bound state}
\label{sec:WeylABS}

We obtain the ABS energies from Eq.\ (\ref{eq:Beenakker}).
The scattering matrix $\hat{s}_{\rm e}$ is random. In the absence of SO interaction, we chose
the random matrix from the circular orthogonal ensemble and disregard its energy-dependence.

We examine the spectrum for many random matrices. About 6\% of them show a gap closing of
the ABS energies indicating the Weyl singularities. The singularities always come in groups of four.
Figure \ref{fig:E0noSO}(a) gives the positions of singularities for a random matrix of choice.
The time-reversal invariance guarantees that the singularity at
$\vec{\varphi}^{(0)} \equiv (\varphi_1^{(0)}, \varphi_2^{(0)}, \varphi_3^{(0)})$ is accompanied by
the singularity of the same topological charge at $-\vec{\varphi}^{(0)}$. Owing to this,
the positions of four singularities and the center of Brillouin zone $(0,0,0)$ are in the same 2D plane.
Figure \ref{fig:E0noSO}(b) shows the lowest positive energy of the ABS in this plane.
The spectrum is symmetric with respect to phase inversion $\vec{\varphi} \to -\vec{\varphi}$.
In the origin, this energy reaches maximum $E = \Delta$. It drops down to two valleys close to
the edges of the Brillouin zone. A zoom into a valley (Fig.\ \ref{fig:E0noSO}(c)) shows that
the energy actually reaches zero in two isolated points.
%Note that the bands of ABS energies are not periodic in the 2D plane.

In Figs.\ \ref{fig:E0noSO}(d) and (e), we show the ABS energies versus $\varphi_1$ at fixed $\varphi_{2,3}$.
In Fig.\ \ref{fig:E0noSO}(d), the choice of $\varphi_{2,3}$ is such that a singularity is reached at some $\varphi_1$.
In Fig.\ \ref{fig:E0noSO}(e), the line spanned by changing $\varphi_1$ passes close to a singularity.
We see the spectrum is conical near the singularity. For $N_i = 1$ in all four leads, they are four positive
and four negative ABS energies. Since the SO interaction is absent, the energies of ABS are doubly degenerate.
In Figs.\ \ref{fig:E0noSO}(d) and (e), the second ABS band is close to $\pm \Delta$.

\begin{figure}
\includegraphics[width=85mm]{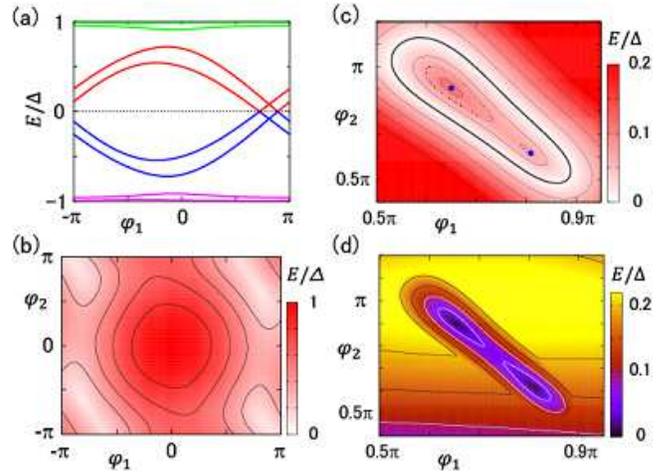}
\caption{(Color online)
Energies of the ABS in the same junction with SO interaction ($p_{\rm SO} = 0.3$).
(a) The ABS energies at $\varphi_2 \simeq 0.62\pi$ and $\varphi_3 \simeq -0.38\pi$ versus $\varphi_1$
(c.\ f.\ Fig.\ \ref{fig:E0noSO}(d)). All the levels are split in spin. The levels pass the position of the singularity at
this choice. The singularity is shifted from zero energy while the spectrum remains conical at this point.
The levels cross zero energy at both side of singularity.
(b) The gray scale plot of the lowest positive ABS energy in the 2D plane that includes the singularities.
Contour lines are the same in Fig.\ \ref{fig:E0noSO}(b), and the overall energy landscape does not change significantly.
The qualitative change in the zoom view of (c). Thick solid line indicates zero energy.
The dots indicate the singularities which are now local maximums of the lowest positive ABS energy.
Thin solid contour lines give $E/\Delta$ in multiples of $0.05$. The dashed line indicates $E/\Delta = 0.075$. 
(d) The gray scale plot of the energy difference between the second and the lowest positive ABS energies.
The contour lines indicate $E/\Delta$ in multiples of $0.05$. The energy difference reaches zero at the positions of
singularities.
}
\label{fig:E0withSO}
\end{figure}

At the same choice of random scattering matrix, we increase the parameter $p_{\rm SO}$ thereby together
continuously increasing the strength of SO interaction. As we see in Fig.\ \ref{fig:E0withSO}(a),
the SO interaction splits the ABS energies in spin. The absence of time-reversibility required for such splitting
comes about non zero $\varphi_{1,2,3}$. As in Fig.\ \ref{fig:E0noSO}, we plot in Figs.\ \ref{fig:E0withSO}(b) and (c)
the lowest positive energy of the Andreev states in the plane that passes through the Weyl singularities.
We see in Fig.\ \ref{fig:E0withSO}(b) that overall energy landscape have not changed significantly in comparison with
Fig.\ \ref{fig:E0noSO}(b). However, as seen in Fig.\ \ref{fig:E0withSO}(c), the landscape has changed drastically in
the vicinity of the singularities. The gap is closed at the closed contour encircling the singularities.
The singularities are shifted to non zero energy while the spectrum remains conical in the vicinity of a singularity
(Fig.\ \ref{fig:E0withSO}(a)). In Fig.\ \ref{fig:E0withSO}(d), we plot energy difference between the second and
the lowest positive energy levels and observed that it goes to zero at the position of the singularities.

\begin{figure}
\includegraphics[width=85mm]{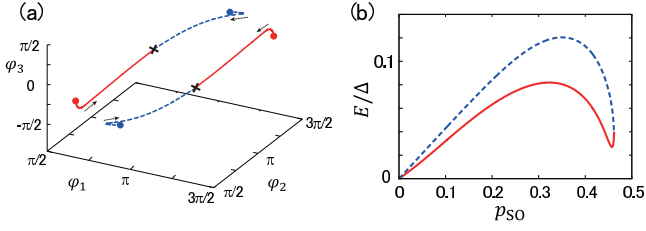}
\caption{(Color online)
(a) The trajectories of the positions of Weyl singularities in a 3D space of the phases
upon increasing the SO strength from $p_{\rm SO} = 0$ to $p_{\rm SO} \simeq 0.462$
when the singularities of the opposite topological charge come together and annihilate.
Solid (dashed) curves give the position of the singularities with positive (negative) charge.
%Note that the space of phases is expanded to see trajectories continuously.
(b) Energies of the singularities with positive (solid) negative (dashed) topological charge versus $p_{\rm SO}$.
The energies merge in the annihilation points.
}
\label{fig:3DWeyl}
\end{figure}

The SO interaction changes the position of the Weyl singularities while preserving their topological charge and
the conical dispersion. When $p_{\rm SO}$ increases gradually, the four Weyl singularities move in the 3D space of
the phases. Figure \ref{fig:3DWeyl}(a) shows trajectories of their positions for $p_{\rm SO}$ arranging from
$p_{\rm SO} = 0$ to $p_{\rm SO} \simeq 0.462$.
%A tuning parameter $p_{\rm SO}$ control a strength of the SO interaction.
%In addition, the parameter modulates a transport property of electron in the scattering matrix continuously.
Solid and dashed curves give the position of the singularities with positive and negative charge. For this particular
choice of the scattering matrix, the singularities of the opposite charge get close to each other upon increasing
the SO strength and eventually annihilate at $p_{\rm SO} \simeq 0.462$, so the junction is not topological
any more (Fig.\ \ref{fig:3DWeyl}(a)).

We compute the energy of the singularity and the plots result in Fig.\ \ref{fig:3DWeyl}(b). We see that this energy
is zero in the absence of SO interaction. The energies of the singularities of different topological charge increase and
become different with increasing $p_{\rm SO}$. At $p_{\rm SO} \gtrsim 0.3$, the energies come close together merging at
the annihilation point $p_{\rm SO} \simeq 0,462$.

\subsection{Effective Hamiltonian}
\label{sec:Heff1}

To prove the generality of the above results, we derive an effective Hamiltonian for ABS that
is valid in the vicinity of a singularity. When the SO interaction is absent, the singularity is at $E = 0$.
At the position of singularity, $\vec{\varphi}^{(0)}$, the scattering matrix $S(\vec{\varphi}^{(0)},0)$
should have an eigenvalue $-1$. However, $S$ is a rather special matrix: at $E = 0$ it can be
presented as $S = s^{(0)*} s^{(0)}$ (cf.\ Eq.\ (\ref{eq:ss})). Here we introduce $s^{(0)} \equiv s(\vec{\varphi}^{(0)},0)$.
We assume nothing about $s^{(0)}$ regarding it as an arbitrary unitary matrix.
This implies that the eigenvalues of $S$ come in complex-conjugated pairs.
If $|a \rangle$ is an eigenvector of $S$ with eigenvalue $e^{i\lambda}$,
$|b \rangle \equiv s^{(0)*} |a^* \rangle$ is an eigenvector with eigenvalue $e^{-i\lambda}$.
This implies that the two orthogonal eigenvectors corresponding to eigenvalue $-1$,
\begin{equation}
s^{(0)*} s^{(0)}| a,b \rangle = -|a,b \rangle.
\label{eq:sspoint1}
\end{equation}
To obtain the effective Hamiltonian, we project the matrix in Eq.\ (\ref{eq:Beenakker}) on
to the space spanned by these two eigenvectors. For small deviations of the phases from $\varphi^{(0)}$,
$\delta \vec{\varphi} = \vec{\varphi} - \vec{\varphi}^{(0)} \ll 1$. We expand the scattering matrix as
$s(\vec{\varphi}) = s^{(0)} e^{iX(\delta \vec{\varphi})} \approx s^{(0)} (1+iX)$, where $X$ is
a Hermitian matrix proportional to $\delta \vec{\varphi}$. Up to the first order in $X$ and $E$, we find
\begin{eqnarray}
S (\vec{\varphi}) &\approx & s^{(0)*} s^{(0)} + i \bar{X}; \hspace{3mm}
\bar{X} \equiv s^{(0)*} s^{(0)} X - s^{(0)*} X^* s^{(0)},
\label{eq:Xcorr} \\
e^{i 2\chi} &\approx & -1 +i2E/\Delta.
\end{eqnarray}
With this, the determinant equation (\ref{eq:Beenakker}) can be presented as an eigenvalue equation for
a $2 \times 2$ effective Hamiltonian~\cite{Riwar},
\begin{equation}
E = \hat{H}; \hspace{3mm}
\hat{H} = \Delta \sum_{j=1}^3 X_j \breve{\Sigma}_j =\frac{\Delta}{2}
\left( \begin{array}{cc}
\langle a| \bar{X} |a \rangle & \langle a| \bar{X} |b \rangle \\
\langle b| \bar{X} |a \rangle & \langle b| \bar{X} |b \rangle
\end{array} \right)
\label{eq:HE0noSO}
\end{equation}
where $\breve{\Sigma}_{1,2,3}$ are the Pauli matrices in the basis of $|a \rangle$ and $|b \rangle$.
Using the relation $|b^* \rangle = s^{(0)}|a \rangle$,  we prove that $\langle b|\bar{X}|b \rangle= - \langle a|\bar{X}|a \rangle$
and real parameters $X_{1,2,3}$ are given simply by $X_1 +iX_2 = -2\langle b|X|a \rangle$ and
$X_3 = -\langle a|X|a \rangle +\langle b|X|b \rangle$. The Hamiltonian in Eq.\ (\ref{eq:HE0noSO})
is a case of Weyl Hamiltonian.

It has opposite eigenvalues, $E = \pm \Delta \sqrt{X_1^2 + X_2^2 + X_3^2}$.
Expanding in $\delta \varphi$, $X_i = X^i_m \delta \varphi_m$, we obtain
\begin{equation}
E = \pm \Delta \sqrt{\sum_{k,m} \delta \varphi_m M_{mk} \delta \varphi_k}; \hspace{3mm}
M_{mk} = \sum_i X^i_m X^i_k
\end{equation}
For 3D space of the phases, the matrix $M$ is positively defined. We reproduce a conical spectrum of
ABS in the vicinity of singularity. For $N$ dimensional space, the matrix $M$ has $N-3$ zero
eigenvalues. We stay at zero energy if we depart from $\varphi^{(0)}$ in this $N-3$ directions.
So that, the singularities are concentrated at $N-3$ dimensional manifold.

Let us take into account weak SO interaction, expanding
\begin{eqnarray}
s(\vec{\varphi}^{(0)}) & = & s^{(0)} e^{i \sum_\alpha \hat{\sigma}_\alpha K_\alpha (\vec{\varphi}^{(0)})}
\approx s^{(0)} \left( 1+i \sum_\alpha \hat{\sigma}_\alpha K_\alpha \right) \\
S (\vec{\varphi}^{(0)}) & = & -\hat{g} s^* (\vec{\varphi}^{(0)}) \hat{g} s(\vec{\varphi}^{(0)})
\approx s^{(0)*} s^{(0)} + i \sum_\alpha \hat{\sigma}_\alpha \bar{K}_\alpha
%; \hspace{3mm}
; \nonumber \\
&& \hspace{10mm}
\bar{K}_\alpha \equiv s^{(0)*} s^{(0)} K_\alpha + s^{(0)*} K_\alpha^* s^{(0)}
\label{eq:Kcorr}
\end{eqnarray}
with $\hat{\sigma}_\alpha$ being Pauli matrices in spin space, $K_\alpha$ being associated Hermitian
matrices in channel space. We project on four dimensional space of spins and vectors $|a,b \rangle$.
We observe that the structure of matrices $\bar{K}_\alpha$ is quit difference from $\bar{X}$.
Using the relations between $|a \rangle$ and $|b \rangle$, we prove $\langle a| \bar{K}_\alpha |b \rangle = 0$ and
$\langle a| \bar{K}_\alpha |a \rangle = \langle b| \bar{K}_\alpha |b \rangle =
- \langle a| K_\alpha |a \rangle + \langle b| K_\alpha |b \rangle \equiv \bm{K}_0$.
With this, the effective Hamiltonian becomes:
\begin{equation}
\hat{H} = \Delta \sum_{j=1}^3 X_j \breve{\Sigma}_j
+ \Delta \hat{\bm{\sigma}} \cdot \bm{K}_0.
\label{eq:HE0}
\end{equation}
The BdG symmetry thus guarantees a special structure of this Hamiltonian
where spin and orbital degree of freedom are totally separated.
$\bm{K}_0$ plays a role of an effective Zeeman field that splits the original conical spectrum.
\begin{equation}
E/\Delta = \sigma_K |\bm{K}_0| \pm \sqrt{\sum_{k,m} \delta \varphi_m M_{mk} \delta \varphi_k}
\end{equation}
with $\sigma_K = \pm$ being the spin projection on the axis of the effective Zeeman field.
The SO interaction does not remove the conical point but rather shift it in energy by $\pm \Delta |\bm{K}_0|$.
In a 3D space, the zero energy is achieved at 2D surface of ellipsoid defined by
$\sum_{k,m} \delta \varphi_m M_{mk} \delta \varphi_k = |\bm{K}_0|^2$.
The singularity is enclosed by the surface. This consideration shows generality of our numerical results.
The size of the ellipsoid enclosing a singularity increases with increasing SO interaction. In Fig.\ \ref{fig:E0withSO}(c),
we see that the ellipsoids enclosing each singularity have already merged together at $p_{\rm SO} = 0.3$.

\section{GAP EDGE TOUCHING}

In this Section, we consider the ABS in the vicinity of the superconducting gap edge $E \approx \Delta$.
We show that the ABS energies reach the GET in the absence of SO interaction at
a 2D surface in the 3D space of the phases. The SO interaction lifts the GET almost
everywhere except particular manifolds: symmetry lines and isolated points.
We investigate these cases separately and establish effective Hamiltonians.

\subsection{GET at symmetry lines}
\label{sec:GET0}

To understand the peculiarities of the GET for multi-terminal junction, let us first consider
two-terminal one with unequal number channels in the left and right lead, $N_{\rm L} < N_{\rm R}$.
The estimation of a number of localized Andreev states is somehow ambiguous. From one hand, one may argue
that there are only $N_{\rm L}$ such state because only that many states are sensitive to the superconducting
phase difference between two leads. From the other hand, the full number of Andreev states is given by
$(N_{\rm L} + N_{\rm R})/2$ provided the total number of channels is even.
The remaining $(N_{\rm R} - N_{\rm L})/2$ states are pinned to the gap edge with no regard for superconducting phase
difference. In a two-terminal junction, these states are indistinguishable from the states of the continuous spectrum.
This is not a case of multi-terminal junction. We note that a multi-terminal junction in fact becomes two-terminal one
if the leads are separated in two groups with superconducting phases are the same within each group.
For instance, in our four-terminal set up, one can choose $\varphi_0 = 0$ and $\varphi_1 = \varphi_2 = \varphi_3 \ne 0$.
Such setting defines a {\it symmetry line} in multi-dimensional space of the phases.
For our example with one channel in each lead, we find extra state pinned at the gap edge along the symmetry line.
In distinction for a two-terminal case, this extra state can not be attributed to the continuous spectrum since
it departs from the gap edge if we go off the symmetry line.

Since the SO interaction does not work for a short two-terminal junction~\cite{Beri},
the SO splitting is also absent at these lines. We postpone the discussion of the details of the spectrum in
the vicinity of symmetry lines to Sec.\ \ref{sec:GETsl} and concentrate now on general situation.

\subsection{ABS energies near the gap edge: general}
\label{sec:GETgene}

Let us consider GET at general position in 3D space of phases. Let us note
mathematical analogy between the spectrum at $E = 0$ and $E = \Delta$. The spectrum is
determined by properties of the scattering matrix $S$ in Eq.\ (\ref{eq:Beenakker}).
Zero energies ($E = 0$) correspond to the eigenvalue of $-1$ of the matrix while GET
correspond to the eigenvalue of $1$. From the other hand, as we have seen in Section \ref{sec:model},
in the absence and presence of SO interaction, we can represent as $S = v^* v$ and $S = - v^* v$,
respectively, $v$ being a general unitary matrix. This establishes a rather unexpected mapping of
the spectrum at $E = 0$ in the presence (absence) of the SO interaction to the spectrum
at $E = \Delta$ in the absence (presence) of the SO interaction.
We have derived in Section \ref{sec:WeylABS} that the ABS reach $E = 0$ at 0D or 2D manifolds in
the 3D spaces in the absence or presence of SO interaction, respectively.
This implies that the GET occurs at 2D and 0D manifolds in the absence and presence of
SO interaction, respectively.

\begin{figure}
\includegraphics[width=85mm]{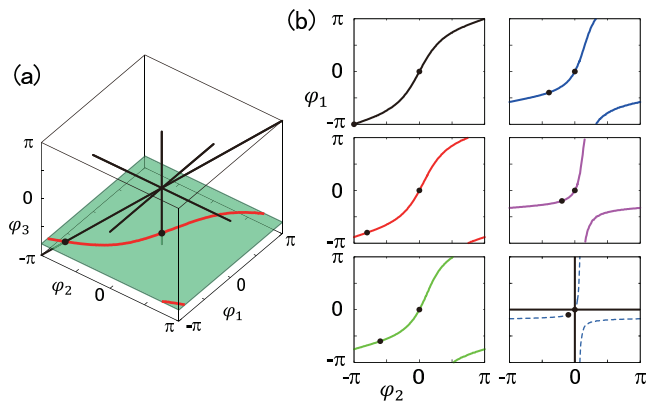}
\caption{(Color online)
GET in a four-terminal junction in the absence of SO interaction.
(a) Curve of the GET at fixed $\varphi_3 = -0.8 \pi$ in the 3D space of the phases.
Four solid lines are symmetry lines, where three of the four superconducting phases are equal.
(b) Curves of the GET at $\varphi_3 = -\pi$ (left upper), $-0.8\pi$ (left middle),
$-0.6\pi$ (left bottom), $-0.4\pi$ (right upper), $-0.2\pi$ (right middle), and $0$ (right bottom panels).
Dashed curves in right bottom panel indicate $\varphi_3 = -0.1\pi$. The black dots give intersection points of
the symmetry lines with the corresponding $\varphi_1$-$\varphi_2$ planes.
}
\label{fig:2Dsurface}
\end{figure}

Let us see this in numerical results.
In the absence of SO interaction, scattering matrix $\hat{s}_{\rm e}$ is chosen randomly from
the circular orthogonal ensemble. Its energy-dependence is disregarded. We examine the spectrum for
many random scattering matrices. The GET is observed for all the matrices.
For a 3D space of the phases, the ABS energy reaches $E = \Delta$ at a 2D surface.
This is illustrated in Fig.\ \ref{fig:2Dsurface}. We fix $\varphi_3$ and find the GET points forming a 1D curve.
Interestingly, the curve passes symmetry lines where we indeed expect GET.
For instance, in Fig.\ \ref{fig:2Dsurface}(a), the curve passes the symmetry line
$\varphi_1 = \varphi_2 = \varphi_3$ at $(\varphi_1 ,\varphi_2) \approx (-\pi , -\pi)$ and passes
the symmetry line $\varphi_1 = \varphi_2 = 0 (= \varphi_0)$ at $(\varphi_1 ,\varphi_2) = (0,0)$.
As $\varphi_3$ is tuned from $-\pi$ to $0$ (Fig.\ \ref{fig:2Dsurface}(b)), the positions of the first passing is
gradually sifted to $(\varphi_1 ,\varphi_2) \approx (0,0)$. At $\varphi_3 = 0$, the curve approaches
two symmetry lines in this plane, $\varphi_1 = \varphi_3 = 0$ and $\varphi_2 = \varphi_3 = 0$.
For positive $\varphi_3$, the spectrum is obtained from the inversion symmetry.
We see that the GET occurs at a single connected 2D surface that includes all four symmetry lines.
In principle, there is nothing to for bit more sophisticated topology of the surface.
However, in a dozen of samples we have explored, we have found no complex topology.

\begin{figure}
\includegraphics[width=85mm]{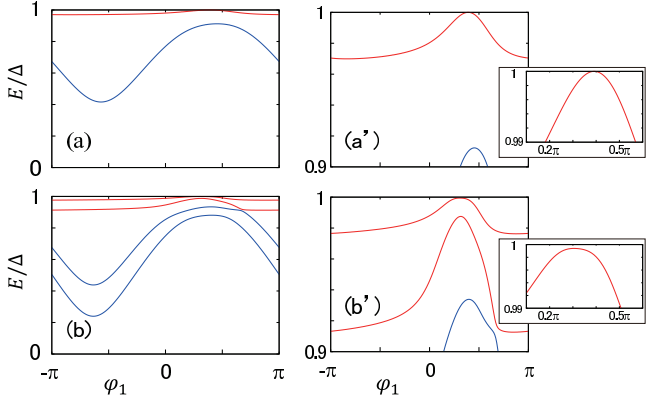}
\caption{(Color online)
Energies of the ABS in a four-terminal junction. The sample is the same as that in Fig.\ \ref{fig:2Dsurface}.
(a) ABS energies versus $\varphi_1$ when spin-orbit interaction is absent. $\varphi_2$ and
$\varphi_3$ are fixed at $0.3\pi$ and $0.5\pi$, respectively. The GET is
obtained at $\varphi_1 \simeq 0.391\pi$. Panel (a') and its inset are enlarged views.
(b) ABS energies when spin-orbit interaction is present ($p_{\rm SO} = 0.5$). The other
parameters and scales of axes are the same as those in upper panels.
}
\label{fig:GETpoint}
\end{figure}

Figures \ref{fig:GETpoint}(a) and (a') show the ABS energies with the GET in the absence of SO interaction.
The energies are plotted versus $\varphi_1$ while $\varphi_2$ and $\varphi_3$ are fixed.
The GET is found at $\varphi_1 \simeq 0.391\pi$. In the vicinity of the touching point,
the spectrum of ABS energy is parabolic rather than conical.
A mirror symmetry of the ABS energies guarantees the GET at $E = -\Delta$ at the same position in
the 3D space of the phases.The SO interaction (Figs.\ (b) \ref{fig:GETpoint}and (b')) lifts
a degeneracy of the energies in spin. In Fig.\ \ref{fig:GETpoint}(b'), one spin-resolved ABS energy comes
very close to the gap edge. However, as seen in the inset, the touching does not take place in
the presence of SO interaction conform to our expectation.

\begin{figure}
\includegraphics[width=85mm]{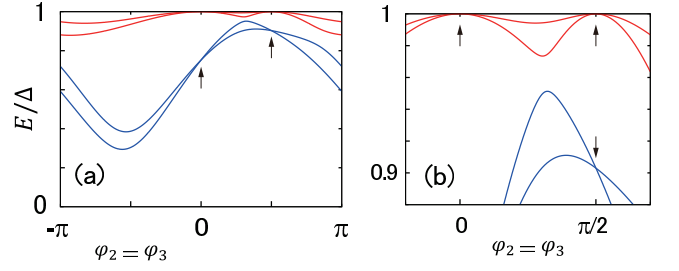}
\caption{(Color online)
Energies of the ABS in the vicinity of symmetry lines. The sample is the same as that in Fig.\ \ref{fig:2Dsurface}.
(a) ABS energies on a line $\varphi_2 = \varphi_3$ at $\varphi_1 = \pi /2$.
A strength of spin-orbit interaction is $p_{\rm SO} = 0.5$.
(b) Enlarged view of panel (a). Arrows indicate points at the symmetry lines.
}
\label{fig:Sline}
\end{figure}

In Fig.\ \ref{fig:Sline}, we illustrate the situation in the vicinity of symmetry lines.
We plot the ABS energies along the line $\varphi_1 = \pi /2$. This line crosses the symmetry lines.
The crossing points are indicated by arrows in the Figure. We see that at these points,
$\varphi_2 = \varphi_3 = 0, \pi/2$, the GET survives in the presence of SO interaction.
Also, the spin splitting of the lowest ABS vanishes at this point.
Then, this proves that the SO interaction is irrelevant at the symmetry lines.

\subsection{effective Hamiltonian in general case}
\label{sec:Heff2}

To prove the generality our numerical results concerning the GET, we derive
an effective Hamiltonian for ABS near $E = \Delta$ taking into account weak SO interaction and
energy-dependence of the scattering matrix. We assume no vicinity of symmetry lines.
This specific situation will be considered separately in the next Section.

Let us first neglect SO interaction and energy-dependence of the scattering matrix.
For a GET point at $\vec{\varphi}^{(0)}$, the matrix $S$ has to have an eigenvalue $1$.
We put $s (\vec{\varphi}^{(0)}) = s^{(0)}$. This eigenvalue is double degenerate;
if $|a \rangle$ is an eigenvector of $S (\vec{\varphi}^{(0)}) = s^{(0)*} s^{(0)}$ with the eigenvalue $1$,
$|b \rangle = s^{(0)} |a^* \rangle$ is also
an eigenvector of $S (\vec{\varphi}^{(0)})$,
\begin{equation}
s^{(0)*} s^{(0)} |a ,b \rangle = |a ,b \rangle.
\label{eq:sspoint2}
\end{equation}
We project the matrix in Eq.\ (\ref{eq:Beenakker}) on to together the space spanned by these two eigenvectors.
Next we expand full scattering matrix in Eq.\ (\ref{eq:Beenakker}) with respect to small phase deviation from
$\varphi^{(0)}$, weak SO interaction, and weak energy-dependence.
%$X(\delta \vec{\varphi})$
%$\hat{\sigma}_\alpha K_\alpha (\vec{\varphi}^{(0)})$
%$E d(\vec{\varphi}^{(0)}) \approx \Delta d(\vec{\varphi}^{(0)}) \equiv D(\vec{\varphi}^{(0)})$.
The first order expansion in $X$ (deviation) and $K_\alpha$ (SO interaction) is the same as given by in
Eqs.\ (\ref{eq:Xcorr}) and (\ref{eq:Kcorr}), respectively.
The energy-dependence is expanded with a Hermitian matrix $d$, 
\begin{eqnarray}
s^{(0)} (E) & = & s^{(0)} e^{iE d(\vec{\varphi}^{(0)})} \approx s^{(0)} (1+iE d) \\
S (\vec{\varphi}^{(0)}, \Delta ) & = & s^{(0)*} (-\Delta ) s^{(0)} (\Delta )
\approx s^{(0)*} s^{(0)} + i \bar{D}
%; \hspace{3mm}
; \nonumber \\
&& \hspace{10mm}
\bar{D} \equiv \Delta (s^{(0)*} s^{(0)} d + s^{(0)*} d^* s^{(0)}).
\label{eq:Dcorr}
\end{eqnarray}
The causality of scattering matrix $s(E)$ implies that all eigenvalue of $d$ are positive.

The first order correction terms are similar to those for Weyl singularity at $E \approx 0$.
However, since the sign of the eigenvalue is opposite, the selection rules for matrix elements
representing the deviation and SO interaction are interchanged. For small phase deviations,
$\langle a |\bar{X}|b \rangle =0$;
$\langle a |\bar{X}|a \rangle = -\langle b |\bar{X}|b \rangle
= \langle a |X|a \rangle - \langle b |X|b \rangle \equiv X_3$.
The expansion of the weak SO interaction gives three independent vectors in spin space,
$\bm{K}_0 \equiv (\langle a |\bm{K}|a \rangle + \langle b |\bm{K}|b \rangle)
= \langle a |\bar{\bm{K}}|a \rangle = \langle b |\bar{\bm{K}}|b \rangle$ and 
$\bm{K}_1 + i\bm{K}_2 \equiv 2 \langle b|\bm{K}|a \rangle = \langle b |\bar{\bm{K}}|a \rangle$.
For weak energy-dependence, $\langle a |\bar{D}|a \rangle = \langle b |\bar{D}|b \rangle
= \langle a |D|a \rangle + \langle b |D|b \rangle \equiv D_0$.
No restriction applies to the off-diagonal matrix elements. So we define
$D_1 + iD_2 \equiv 2\langle b |D|a \rangle = \langle b |\bar{D}|a \rangle$.
The positivity of $d$ implies $D_0 \geq \sqrt{D_1^2 + D_2^2}$.

To deviate an effective Hamiltonian, we introduce the energy deviation from the edge $\epsilon = 1-|E|/\Delta$.
Up to the first order in $\sqrt{\epsilon}$, we obtain $e^{i 2\chi} \approx 1\pm i\sqrt{8\epsilon}$.
Summarizing all these, we obtain that the spectrum is defined by the following eigenvalue equation
\begin{equation}
\hat{H} = \sqrt{2\epsilon}; \hspace{3mm}
%\label{eq:GETeq}
\hat{H} = X_3 \breve{\Sigma}_3
+ \sum_{j=0}^2 \hat{\bm{\sigma}} \cdot \bm{K}_j \breve{\Sigma}_j
+ \sum_{j=0}^2 D_j \breve{\Sigma}_j.
\label{eq:HED}
\end{equation}
Here $\breve{\Sigma}_0 \equiv \breve{1}$. The full analysis of this Hamiltonian is involved.
Here, we discuss several simple cases. Let us first neglect both SO interaction and energy-dependence.
The eigenvalues of the Hamiltonian are in this case simply $\pm |X_3|$. The negative eigenvalue does not
lead to any localized state. The positive eigenvalue provides $\epsilon = X_3^2/2$ since $X_3$ is linear in
phase deviation this defines a parabolic spectrum touching the gap edge at $X_3 = 0$.
We see that the GET requires one condition $X_3 = 0$ to be fulfilled in distinction from
three condition $X_j = 0$ required for Weyl singularity. This is why the manifold of GET point in
$N$-dimensional space of phases generally has dimension $N-1$ in the absence of SO interaction and
energy-dependence of scattering matrix.

\begin{figure}
\includegraphics[width=85mm]{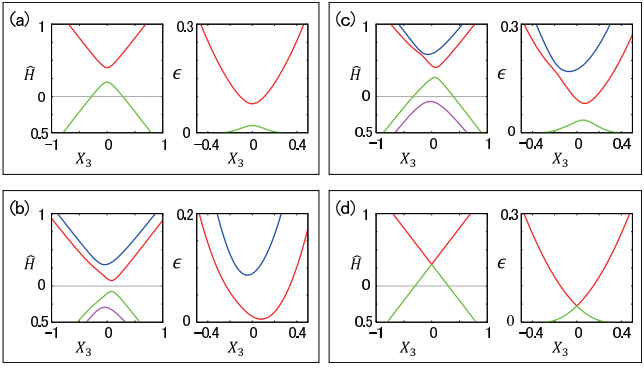}
\caption{(Color online)
Fine structure of GET. $X_3$ gives the distance from the GET.
(a) The effect of energy-dependence. We set $D_0 = 0.3$ and $\sqrt{D_1^2 + D_2^2} = 0.1$.
(b) The effect of SO interaction. $\bm{K}_0$ is parallel to $\bm{e}_3$.
We set $K_0 = 0.1$ and $\sqrt{|\bm{K}_1|^2 + |\bm{K}_2|^2} = 0.2$.
Directions of the vectors are chosen randomly.
(c) The combined effect of the energy-dependence and SO interaction. $D_j$ and $\bm{K}_j$
are the same as those in (a) and (b), respectively.
(d) Weyl singularity near the gap edge. We put $X_1 = X_2 = 0$ and $D_0 = 0.3$.
}
\label{fig:spctGET}
\end{figure}

If we take energy-dependence into account, the eigenvalues are given by $D_0 \pm \sqrt{X_3^2 + D_1^2 + D_2^2}$.
The plot of eigenvalues and spectrum is given in Fig.\ \ref{fig:spctGET}(a).
We see that the energy-dependence modifies the GET in a rather complex way.
The spectrum in the vicinity of the GET line forms two bands. While one of the band never touches the edge,
the other band exist only in the vicinity of the line and merges with the continue upon increasing $X_3$.

Let us neglect now the energy-dependence and take into account SO interaction.
When $\bm{K}_j$ terms are taken into account, the four eigenvalues of $\hat{H}$ are
\begin{equation}
\pm \sqrt{ X_3^2 + L \pm 2\sqrt{ (\bm{K}_0 \cdot \bm{K}_1)^2 + (\bm{K}_0 \cdot \bm{K}_2)^2
+ | X_3 \bm{K}_0 - \bm{K}_1 \times \bm{K}_2 |^2 } }
\label{eq:eigenGETK}
\end{equation}
with $L = |\bm{K}_0|^2 + |\bm{K}_1|^2 + |\bm{K}_2|^2$.
Two positive eigenvalues define two spin-split bands that never touch the edge (Fig.\ \ref{fig:spctGET}(b))
unless a special condition discussed below is fulfilled.
%This condition is rather involved and is formulated in the Appendix.
The minimal energy distance to the gap edge is not achieved at $X_3 = 0$ but rather is shifted depending on
the parameters of the SO interaction.

If the energy-dependent terms and the spin-splitting are the same order of magnitude, the picture of the GET
can become complex. For instance for an example given in Fig.\ \ref{fig:spctGET}(c), there are only three bands.
One existing only in the vicinity of the gap edge line.

Let us now discuss a special condition of the GET in the presence of sufficiently strong SO interaction.
To analyze this, let us derive an effective Hamiltonian with strong SO interaction assuming a GET point to
be present at $\vec{\varphi}^{(0)}$. Strong SO interaction guarantees that only two spin-dependent eigenvectors
are important instead of the four as in the previous consideration.
%Spin-resolved two states $|a \uparrow \rangle$ and $|b \downarrow \rangle$ are taken into account.
The two eigenvectors $|a \rangle$ and $|b \rangle = u^{(0)} |a^* \rangle$ satisfy
\begin{equation}
- u^{(0)*} u^{(0)} |a,b \rangle = |a,b \rangle
\label{eq:uupoint}
\end{equation}
with $u^{(0)} \equiv u (\vec{\varphi}^{(0)})$. The problem is thus mathematically equivalent to our consideration of
Weyl singularity in Sec.\ \ref{sec:Heff1}. The first order expansion in $X$ (deviation) and $\Delta d$
(energy-dependence) results in correction terms $\bar{X} \equiv - u^{(0)*} u^{(0)} X + u^{(0)*} X^* u^{(0)}$ and
$\bar{D} \equiv -\Delta( u^{(0)*} u^{(0)} d + u^{(0)*} d^* u^{(0)})$, respectively.
We prove the selection rules for elements of matrices $\bar{X}$ and $\bar{D}$; $\langle a|\bar{D}|b \rangle = 0$.
The effective Hamiltonian thus reads
\begin{equation}
\hat{H} = \sum_{j=1}^3 X_j \breve{\Sigma}_j + D_0 \breve{\Sigma}_0.
\label{eq:HEDstrong}
\end{equation}
with $X_1 +i X_2 \equiv \langle b|\bar{X}|a \rangle = 2 \langle b|X|a \rangle$,
$X_3 \equiv \langle a|\bar{X}|a \rangle = - \langle b|\bar{X}|b \rangle = \langle a|X|a \rangle - \langle b|X|b \rangle$, and
$D_0 \equiv \langle a|\bar{D}|a \rangle = \langle b|\bar{D}|b \rangle = \langle a|D|a \rangle + \langle b|D|b \rangle \geq 0$.
We keep here energy-dependent term $\propto D_0$ that is absent in the Hamiltonian (\ref{eq:HE0noSO}).
We have just found a Weyl singularity with a conical point near the gap edge. The conical point requires
three conditions to be fulfilled $X_j = 0$. Therefore we expect the points to form $N-3$ dimensional manifold
in the $N$ dimensional spaces. The energy-dependent term shifts the energy of the conical point from
the gap edge similar to the effect of the SO interaction in Eq.\ (\ref{eq:HE0}) (Fig.\ \ref{fig:spctGET}(d)).
We have tried to find this singularities in our numerical simulation. So far, we have found none.
The reason of this is not completely clear for us. We hypothesize that probability to find a random scattering
matrix with such singularity is low because the GET point in the presence of SO interaction
tend to stick to the symmetry line.

\subsection{The vicinity of a symmetry line}
\label{sec:GETsl}

Let us consider an effective Hamiltonian in the vicinity of symmetry lines. The SO interaction is assumed to be strong.
We concentrate on a four-terminal junction. At the symmetry lines, three of the four superconducting phases are equal.
The electron scattering matrix can be presented in a block structure,
\begin{equation}
\hat{s}_{\rm e} =
\left( \begin{array}{c|c}
r_1                               & \hspace{10pt} t_{13} \hspace{5pt} \\ \hline
\lower1ex\hbox{$t_{31}$} & \lower1ex\hbox{$r_3$} \lower3ex\hbox{}
\end{array} \right).
\end{equation}
Assuming there are $N_3$ channels in three leads having the same phase and $N_1$ channels in the other lead.
$r_3$ is a $2N_3 \times 2N_3$ reflection matrix for three leads having the same phase.
$r_1$ is that of the other lead, $t_{13}$ and $t_{31}$ are the transmission matrices between the three leads and
the other lead, their dimensions are $2N_1 \times 2N_3$, $2N_3 \times 2N_1$, correspondingly.
At the symmetry line, we have $2(N_3 - N_1)$ independent vectors satisfying $t_{13}|\psi \rangle = 0$.
These states are disconnected from the other lead. Their energies are precisely at the gap edge,
therefore
\begin{equation}
- \hat{g} s^{(0)*} \hat{g} s^{(0)}
\left( \begin{array}{c}
0 \\
|\psi \rangle
\end{array} \right)
=
\left( \begin{array}{c}
0 \\
|\psi \rangle
\end{array} \right)
\end{equation}
is satisfied. Similar to previous considerations, these eigenvectors can be arranged into
$N_3 - N_1$ conjugated pairs: if $|a \rangle$ is an eigenvector, $|b \rangle \equiv \hat{g} r_3^* |a^* \rangle$
is also an orthogonal eigenvector. Note that $r_3$ is a unitary matrix.

\begin{figure}
\includegraphics[width=85mm]{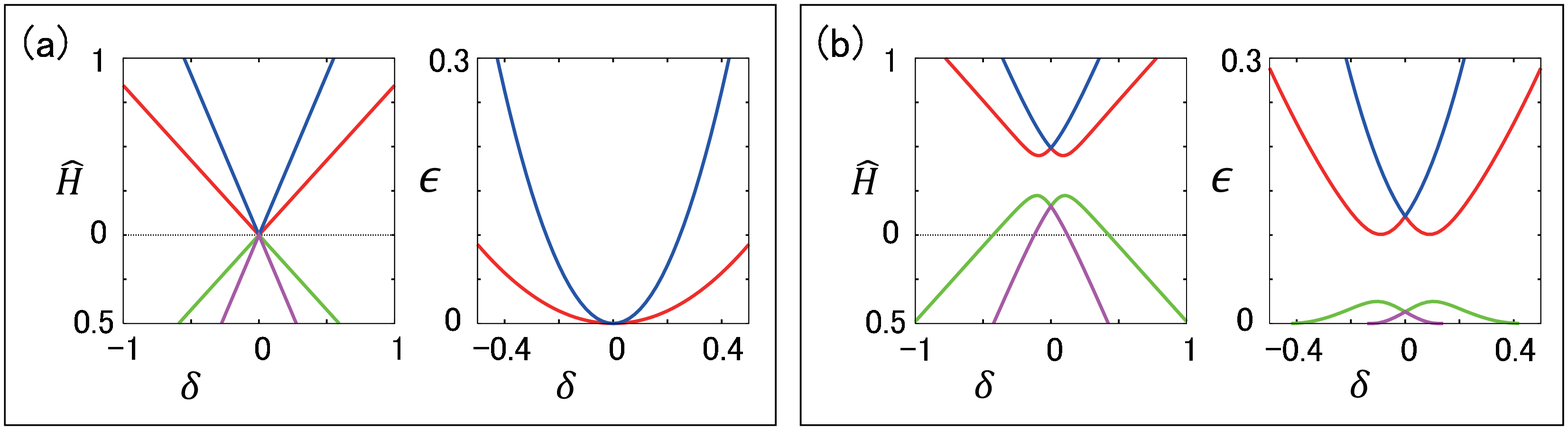}
\caption{(Color online)
Fine structure of GET in the vicinity of a symmetry line with strong SO interaction.
$\delta \equiv |\bm{h}_{ab}| = |\bm{h}_{\tilde{a}\tilde{b}}| = \sqrt{h_0^2 + |\bm{h}|^2}$ is
linear to a phases deviation from the line.
(a) Neglecting energy-dependence.
(b) The effect of energy-dependence. We set $\sqrt{ D_{ab} D_{\tilde{a} \tilde{b}} } = 0.3$
and $\sqrt{ D_0^2 + |\bm{D}|^2} = 0.1$.
}
\label{fig:spctSL}
\end{figure}

To derive the effective Hamiltonian, we project Eq.\ (\ref{eq:Beenakker}) onto the subspace of these eigenvectors.
The consideration can be done for arbitrary dimension, but for the sake of comprehensibility,
we concentrate on this situation with a single channel in each lead. In this case, $N_3 =3$ and $N_1 = 1$ and
we project on four orthogonal states $|a \rangle ,|b \rangle ,|\tilde{a} \rangle, |\tilde{b} \rangle$.
For a small phase deviation from the symmetry line, $S(\vec{\varphi}) \approx S(\vec{\varphi}^{(0)}) + \delta S$.
The matrix element $(\delta S)_{\alpha \beta}$ can be represented as
\begin{equation}
(0, \langle \alpha |)^{\rm T}
\delta S
\left( \begin{array}{c}
0 \\
|\beta \rangle
\end{array} \right)
=i \langle \alpha | r_3^\dagger r_3 \hat{\delta} + \hat{\delta} r_3^\dagger r_3
- 2 r_3^\dagger \hat{\delta} r_3 | \beta \rangle
\equiv i 2 \bar{X}_{\alpha \beta}.
\end{equation}
Here, $\hat{\delta}$ is the diagonal matrix of the phase deviations.
Note that we use $\langle \alpha | t_{13}^\dagger t_{13} | \beta \rangle = 0$ to arrive at this.
A matrix representation of this correction term $\bar{X}_{\alpha \beta}$ for the four bases gives
a Hamiltonian in the vicinity of the symmetry line
\begin{equation}
\hat{H} = \bar{X} =
\left( \begin{array}{cc}
\hat{\bm{\sigma}} \cdot \bm{h}_{ab}     & -i h_0 + \hat{\bm{\sigma}} \cdot \bm{h} \\
i h_0 + \hat{\bm{\sigma}} \cdot \bm{h} & \hat{\bm{\sigma}} \cdot \bm{h}_{\tilde{a}\tilde{b}}
\end{array} \right).
\end{equation}
Here, we use $r_3^{\rm T} = - \hat{g} r_3 \hat{g}$.
The parameters in block diagonal components are defined as
$h_{ab,1} +i h_{ab,2} \equiv \bar{X}_{ab} = 2 \langle a| \hat{\delta} |b \rangle$,
$h_{ab,3} \equiv \bar{X}_{aa} = - \bar{X}_{bb} = \langle a| \hat{\delta} |a \rangle - \langle b| \hat{\delta} |b \rangle$,
and the same way for $\bm{h}_{\tilde{a}\tilde{b}}$.
Those in off-diagonal components are
$h_{3} + ih_{0} \equiv \bar{X}_{\tilde{a} a} = - \bar{X}_{\tilde{b} b}^*
= \langle \tilde{a}| \hat{\delta} |a \rangle - \langle b| \hat{\delta} |\tilde{b} \rangle$, and
$h_{1} + ih_{2} \equiv \bar{X}_{\tilde{b}a} = \bar{X}_{\tilde{a}b}^* = \langle \tilde{b}| \hat{\delta} |a \rangle$.
The eigenvalues of $\hat{H}$ are
\begin{equation}
\pm \sqrt{ h_0^2 + L^\prime \pm \sqrt{ ( \bm{h}_+ \cdot \bm{h}_- )^2/4 + ( \bm{h} \cdot \bm{h}_+ )^2
+ | h_0 \bm{h}_+ + \bm{h} \times \bm{h}_- |^2 }}
\label{eq:eigenSL}
\end{equation}
with $L^\prime = ( |\bm{h}|^2 + |\bm{h}_{ab}|^2 + |\bm{h}_{\tilde{a}\tilde{b}}|^2 )/2$ and
$\bm{h}_\pm = \bm{h}_{ab} \pm \bm{h}_{\tilde{a}\tilde{b}}$. This is similar to Eq.\ (\ref{eq:eigenGETK}),
however, in the case under consideration, all elements of the effective Hamiltonian are proportional to
the phase deviations. This results in a linear splitting of four-hold degenerate eigenvalue and two bands
touching the gap edge at the symmetry line (Fig.\ \ref{fig:spctSL}(a)).

The energy-dependence is taken into account the same way as Sec.\ \ref{sec:Heff2}.
It gives a correction term in the Hamiltonian that importantly does not vanish at $\hat{\delta} \to 0$,
\begin{equation}
\hat{H} = \bar{X} +
\left( \begin{array}{cc}
D_{ab}                                              & D_0 - i \hat{\bm{\sigma}} \cdot \bm{D} \\
D_0 + i \hat{\bm{\sigma}} \cdot \bm{D} & D_{\tilde{a}\tilde{b}}
\end{array} \right)
\end{equation}
with $D_{ab} = \langle a|D|a \rangle + \langle b|D|b \rangle$ and the same for $D_{\tilde{a}\tilde{b}}$,
$D_{0} + iD_{3} = \langle \tilde{a}|D|a \rangle + \langle b|D|\tilde{b} \rangle$, and
$D_{2} - iD_{1} = \langle b|D|\tilde{a} \rangle - \langle \tilde{b}|D|a \rangle$.
Here, $D_{ab} D_{\tilde{a}\tilde{b}} \geq D_0^2 + |\bm{D}|^2$ due to the causality.
This guarantees the positive eigenvalues at the symmetry lines, $\hat{\delta} = 0$, given by
$(D_{ab} + D_{\tilde{a}\tilde{b}})/2 \pm \sqrt{ (D_{ab} - D_{\tilde{a}\tilde{b}})^2 /4 + D_0^2 + |\bm{D}|^2 }$.
The eigenvalues are doubly degenerate. The deviation from the symmetry line gives rise to
a linear splitting of the eigenvalues at further increase $\hat{\delta}$. The eigenvalues approach
a linear asymptotics given by Eq.\ (\ref{eq:eigenSL}). We illustrate the spectrum in Fig.\ \ref{fig:spctSL}(b).

\section{CONCLUSIONS AND DISCUSSION}

We have studied the singularities and peculiarities in the ABS spectrum of a Josephson junction
connected to $N$ superconducting leads. The ABS energies in such junctions depend on  $N - 1$ independent
superconducting phase differences, $E (\varphi_1, \varphi_2, \cdots , \varphi_{N-1} )$ being a periodic function of all phases.
Therefore, they can be regarded as energy bands in the $N - 1$ dimensional periodic solid,
if one associates $\varphi_j$ with quasimomenta. We have concentrated on the singularities related to
topological properties and use numerical illustrations of the spectrum and as well as derive effective Hamiltonians to
describe the vicinity of the singularities. The illustrations are made for a four-terminal short junction.
In this case, the energies of ABS correspond to the bands in a 3D solid. The ABS energies are calculated from
Beenakker's determinant equation using scattering matrix.

We reveal the  singularities in the vicinity of zero energy and near the gap edge.
We establish a mathematical analogy between the spectrum at $E = 0$ and $E = \Delta$.

First, we have considered Weyl singularities near zero energy.
When the SO interaction is absent, the singularities are found at $E = 0$ accompanying conical spectrum.
The Weyl singularities occur at isolated 0D points in the 3D space of the phases.
The SO interaction splits the singular points to mirror symmetric positive and negative energies in spin.
A small modification of scattering matrix only shifts the position of the Weyl singular points
but does not eliminate those since they are topologically protected.
In the presence of SO interaction, zero energy points in the vicinity of Weyl singularities form a 2D manifold in
the 3D space of the phases. This 2D manifold encircles the singular point.
To prove our numerical results, we have derived an effective Hamiltonian that is valid in the vicinity of
the singularity and at weak SO interaction.
Eigenvalues of the Hamiltonian reproduce a conical singularity in the spectrum.
They reach zero at an ellipsoid enclosing the singular point.

Exploiting  the  mentioned analogy between the ABS at $E = 0$ and $E = \Delta$, we have investigated the spectrum in 
the vicinity of the gap edge. The analogy implies the GET occurs at a 2D surface in the 3D space in the absence of
SO interaction and at isolated  points in the presence of SO interaction.
Thus, the SO interaction generally lifts the GET except specific situations.
We have  established effective Hamiltonians for two specific cases: symmetry lines and isolated points. 
We have also taken into consideration a weak energy dependence of the scattering matrix relevant for fine structure of GET.
The effective Hamiltonians derived prove the generality of our numerical results.
For the GET in a strong SO interaction, the Hamiltonian indicates a Weyl singularity with a conical spectrum.
The singularity is shifted from the edge by the energy-dependent term similar to the effect of SO interaction at $E \approx 0$.
However, we do not find the isolated GET point in our numerical simulations.

At a symmetry line, since three of the four phases are equal, the four-terminal junction can
be regarded as a two-terminal junction with unequal channel numbers in the leads.
Then, only a single ABS is sensitive to the phase differences while the other stick to the gap edge.
We have also derived an effective Hamiltonian in the vicinity of the symmetry lines.
The energy-dependence of the scattering matrix lifts the GET and adds another Andreev state,
which is localized in the vicinity of the symmetry line.

Our study provides a numerical and analytical evidence for the Weyl points in multi-terminal Josephson junctions.
These points are different from Weyl singular points from that predicted and recently found in 3D solids,
such as TaAs~\cite{SuYXu1,Lv,Yang}, TaP~\cite{NXu}, and NbAs~\cite{SuYXu2}.
In the 3D solids, the absence of inversion symmetry in the material and removing the spin degeneracy are
essential for a stable Weyl point. However, multi-terminal Josephson junction shows stable Weyl points
even though the ABS energies are not spin-split.
Our study thus facilitates an alternative method of realization of Weyl fermions in condensed matter.

The Weyl points in multi-terminal Josephson junction should provide positive and negative
topological charges in the space of the phases. The charges are sources of the Berry curvature fields.
Riwar {\it et al.}~\cite{Riwar} have proposed an experimental setup to detect the Chern number by
measuring the quantized transconductance, in similarity with the quantum Hall effect.
In their scheme, the Chern number is defined as the integral of the Berry curvature field on
a 2D plane in the space. Physically, the integration is achieved by sweeping the phases by
applying finite bias voltages to the superconducting leads.
Riwar et al. considered a situation without SO interaction where the Weyl singular points appear
always at E=0 and are doubly degenerate with respect to spin. Then the lowest positive
(and the highest negative) levels of the ABS are relevant to the Chern number.
Near the Weyl point, the quantization of transconductance may be sensitive to the temperature of
the junction owing to the undesired thermal activation of a quasiparticle  in the lowest subband.
In this paper, we take into account the SO interaction and consider Weyl singularities of two types.
Although the SO interaction shifts the conical point from E=0, it does not influence the Chern number
arising from the singularity. Therefore, the topological charge may be still detected by measuring
the quantized transconductance. The same applies to the singularity at the gap edge.
The SO shift of the singular point from the Fermi energy implies that the probability of
undesired thermal activation of a quasiparticle state near the point of the singularity is
small at sufficiently low temperatures, this facilitates the observation of the topological effect.

%Multi-terminal junction has many possibilities as future works. If more than four superconducting
%leads are connected to the junction, its band structure is in more than three dimensional space,
%which is impossible in real materials. Our study indicates topological singularities even for a band
%in 3D square lattice. More attractive physics may be there in high dimensional spaces.

\section*{ACKNOWLEDGMENT}
We appriciate the fruitful discussion with Roman-Pascal Riwar, Manuel Houzet, Julia S.\ Meyer in Universit\'{e} Grenoble Alpes.
This work has been partially supported by JSPS Postdoctoral Fellowships for Research Abroad and
the Nanosciences Foundation in Grenoble, in the frame of its Chair of Excellence program grand in Grenoble.

\end{document}